\documentclass[aps, prd, twocolumn, 10pt, superscriptaddress, nofootinbib]{revtex4}
\usepackage[dvipsnames]{xcolor}
\usepackage{physics}
\usepackage{braket}
\usepackage{scalerel}
\usepackage{blindtext} 
\usepackage{epsfig, cancel}

\usepackage{latexsym}
\usepackage{natbib, comment}
\usepackage{mathrsfs,amsmath,amssymb,amsthm,amsfonts,tikz,graphicx,accents,hyperref,color}
\usepackage{url}
\usepackage{dcolumn}
\usepackage{multirow}
\usepackage{color}
\usepackage{cancel}
\usepackage{soul}
\usepackage[normalem]{ulem}
\usepackage{txfonts}
\usepackage{epsfig}
\usepackage{psfrag}
\usepackage{subfigure}
\hypersetup{colorlinks=true}
\usepackage{mathtools}
\usepackage{enumitem}
\usepackage{float}
\usepackage{caption,ragged2e}
\usetikzlibrary{decorations.markings,fpu}

\usetikzlibrary{calc}

\def\H0{{\text{H}\hspace*{-2.05mm}\text{H} 0\hspace*{-1.35mm}0\ }}

\usepackage[all]{xy}

\usepackage{slashed}
\usepackage{slashed,ccaption}
\usepackage{multirow}
\usetikzlibrary{calc} 

\usepackage{caption}

\newcommand{\beps}{\boldsymbol{\epsilon}}
\usepackage{array}
\hypersetup{ linktoc=all,
    colorlinks, linkcolor={palatinateblue},
    citecolor={red}, urlcolor={amaranth} 
}

\graphicspath{{Images/}}

\renewcommand{\d}[1]{\ensuremath{\operatorname{d}\!{#1}}}

\DeclareSymbolFont{extraup}{U}{zavm}{m}{n}
\DeclareMathSymbol{\varheart}{\mathalpha}{extraup}{86}
\DeclareMathSymbol{\vardiamond}{\mathalpha}{extraup}{87}
\makeatletter
\renewcommand*{\@fnsymbol}[1]{\ensuremath{\ifcase#1\or \clubsuit \or \vardiamond \or \varheart\or
    \spadesuit\or \mathparagraph\or \|\or **\or \dagger\dagger
    \or \ddagger\ddagger \else\@ctrerr\fi}}
\makeatother

\definecolor{rosy}{RGB}{230,235,252}
\definecolor{myframetitle}{RGB}{90,89,170}
\definecolor{myblocktitle}{RGB}{140,185,249}
\definecolor{mytitle}{RGB}{10,80,26}

\definecolor{darkgreen}{RGB}{27,130,45}
\definecolor{darkblue}{rgb}{0,0,0.3}
\definecolor{darkred}{rgb}{0.7,0,0}

\definecolor{light gray}{RGB}{220,220,220}
\definecolor{dark purple}{RGB}{108,0,217}
\definecolor{pink}{RGB}{190,20,100}
\definecolor{orang}{RGB}{193,63,0}
\definecolor{green}{RGB}{11,98,17}
\definecolor{darkpink}{RGB}{153,0,76}
\definecolor{bluegreen}{RGB}{0,102,102}
\definecolor{greenlagan}{RGB}{0,102,0}
\definecolor{redgreen}{RGB}{102,102,0}
\definecolor{Redgreen}{RGB}{153,76,0}
\definecolor{vividviolet}{rgb}{0.62, 0.0, 1.0}
\definecolor{amaranth}{rgb}{0.9, 0.17, 0.31}
\definecolor{palatinateblue}{rgb}{0.15, 0.23, 0.89}
\definecolor{brightpink}{rgb}{1.0, 0.0, 0.5}
\definecolor{cornflowerblue}{rgb}{0.39, 0.58, 0.93}
\definecolor{deepcarminepink}{rgb}{0.94, 0.19, 0.22}
\definecolor{radicalred}{rgb}{1.0, 0.21, 0.37}

\makeatletter \@addtoreset{equation}{section}

\newcommand\ignore[1]{}
\usepackage[most]{tcolorbox}

\tcbset{highlight math style={left=02mm,right=02mm,top=02mm,bottom=02mm}} 
\usepackage{empheq}

\newcommand\inbox[1]{\tcbset{fonttitle=\scriptsize} \tcboxmath[colback=white,colframe=black!70]{#1}}

\begin{document}

\title{Gravitational Entropy And The Second Law of Thermodynamics for Causal Observers}
\author{V.~R.~Shajiee}\email{v.shajiee@ipm.ir}
\affiliation{School of Physics, Institute for Research in Fundamental Sciences (IPM), P.O.Box 19395-5531, Tehran, Iran}

\author{M.~M.~Sheikh-Jabbari}\email{jabbari@theory.ipm.ac.ir}
\affiliation{School of Physics, Institute for Research in Fundamental Sciences (IPM), P.O.Box 19395-5531, Tehran, Iran}

\author{V.~Taghiloo}\email{v.taghiloo@iasbs.ac.ir}
\affiliation{Department of Physics, Institute for Advanced Studies in Basic Sciences (IASBS), P.O.Box 45137-66731, Zanjan, Iran}
\affiliation{School of Physics, Institute for Research in Fundamental Sciences (IPM), P.O.Box 19395-5531, Tehran, Iran}

\begin{abstract}
It is well established that black holes possess entropy and behave as thermodynamic systems. Associating entropy with gravitational fields has not remained limited to black holes, necessitating the notion of the second law of thermodynamics in gravitating systems. There have been many ideas and attempts to prove the second law within gravitating systems starting from first principles. Within the covariant phase space formalism, we define gravitational entropy as the charge associated with the local boosts, detaching the gravitational entropy from horizons or trapped surfaces as well as from the diffeomorphisms as symmetry generators. Using this definition for the Einstein gravity case, we compute variations of the entropy along the path of any causal free-fall observer and establish that the entropy variations are always non-negative if the matter content satisfies the strong energy condition integrated along any segment of the observer's trajectory. 

\end{abstract}
\maketitle

\section{Introduction}\label{sec:introduction}

The modern view on gravity, as formulated through Einstein's equivalence principle and general relativity (GR), relates gravitational interactions to the fabric of spacetime. Consequently, the causal structure of spacetime, how light rays travel, becomes a dynamical notion and is affected by the matter content in the spacetime. This framework naturally yields the profound concept of horizons: typically null hypersurfaces that partition spacetime into causally disconnected regions. Therefore, some parts of spacetime may not be available to a specific set of observers, while other observers can access them. The prime examples of spacetimes with horizons are black hole spacetimes; see, e.g., \cite{Hawking:1973uf, Wald:1984rg, Grumiller:2022qhx}.

The second law of thermodynamics, that the total entropy of an isolated/closed system cannot decrease in time, is one of the most fundamental laws of physics. In the presence of horizons, as famously noted by Bekenstein \cite{Bekenstein:1972tm,  Bekenstein:1973ur}, a set of observers who can access only one side of a horizon record a decrease in the total entropy once a thermodynamic system becomes inaccessible to the observer when passing through their horizon, hence violating the second law. To avoid this, one may associate the entropy with the horizon to compensate for the part of spacetime not available to the set of observers. If the entropy associated with the horizon obeys the Bekenstein-Hawking area law (the entropy associated with the horizon is its area over 4 times the Newton constant in natural units), one obtains a notion of the generalized second law where the entropy of the system plus the part of spacetime bound by the horizon does not decrease in time \cite{Bekenstein:1972tm, Bekenstein:1974ax}.

The Bekenstein-Hawking area law and, more generally, laws of thermodynamics in gravitational systems are mostly discussed when dealing with black holes \cite{Bardeen:1973gs, Wald:1999vt} (in particular black holes with Killing horizons \cite{Wald:1993nt, Iyer:1994ys}); see \cite{Hawking:1973uf, Wald:1984rg, Wald:1999vt, Grumiller:2022qhx} for detailed discussions. However, gravitational entropy is by no means limited to black holes. Horizons can be observer-dependent (e.g., apparent horizons \cite{Hayward:1993wb}) rather than global, intrinsic features of spacetime (like event horizons) \cite{Hawking:1971vc, Jacobson:1993xs}. This prompts the viewpoint that 
gravitational entropy is not just a feature of the black hole horizons, but should be a general feature of any gravitating system and fabric of spacetime. And, notions as fundamental as entropy and the second law should be related/connected to equally fundamental features, like the equivalence principle that manifests itself through diffeomorphism invariance (general covariance). 

The first systematic step in defining the entropy was taken by Wald in his seminal paper \cite{Wald:1993nt}, where the entropy of a black hole (with a bifurcate Killing horizon) was introduced as the Noether charge associated with the Killing vector field generating the horizon of the black hole. This work was completed by deriving the first law for the same class of black holes in a generic diffeomorphism-invariant gravity theory \cite{Iyer:1994ys}. There have been other proposals for defining entropy in gravitating systems from first principles. Some of them are based on the notion of entropy bounds, an upper bound on the entropy that can be accumulated in a region of space(time), like the Bekenstein bound~\cite{Bekenstein:1980jp} and the Bousso covariant entropy conjecture~\cite{Bousso:1999xy} while some follow parallel ideas like those in \cite{Jacobson:1993xs, Jacobson:1993vj}, yielding Jacobson's insightful derivation of GR from thermodynamics \cite{Jacobson:1995ab}.

In this work, we extend and generalize Wald's proposal within the covariant phase space formalism (CPSF) \cite{Lee:1990nz, Barnich:2007bf, Harlow:2014yka, Compere:2018aar, Grumiller:2022qhx}. We define the gravitational entropy of a part of space enclosed in the codimension-2 compact (finite area) spacelike hypersurface $\Sigma$ as the surface charge associated with local boosts along the transverse directions to $\Sigma$; see \cite{Jacobson:2015uqa, Neiman:2013taa} for related ideas. This definition enables us to disentangle the notion of entropy from codimension-1 surfaces like (Killing) horizons \cite{Wald:1993nt} or light sheets \cite{Bousso:1999xy}, as well as disentangling it from diffeomorphisms and vector fields as symmetry generators. Our definition also resolves a possible non-integrability of the Iyer-Wald entropy \cite{Iyer:1994ys}, the charge  associated with the horizon-generating Killing vector field normalized by the surface gravity
\cite{Hajian:2015xlp}.

Any reliable definition of the entropy must exhibit the second law. In the context of gravitational systems, the general idea followed in the literature is that the second law is a consequence of some energy condition that standard matter should obey, usually supplemented with other properties or features like having a Killing or event horizon or a (marginally) trapped surface, e.g. see \cite{Hayward:1993wb, Ashtekar:2002ag, Ashtekar:2003hk, Andersson:2005gq, Booth:2005qc, Andersson:2007gy, Bousso:2015qqa, Bousso:2015mqa}. The first ``proof'' for the second law is Hawking's seminal area theorem \cite{Hawking:1971vc}, which invokes the weak energy condition. In a more restricted setting, the second law was associated with the null energy condition (NEC) \cite{Bardeen:1973gs, Parikh:2015ret}; see \cite{Rignon-Bret:2023fjq} and references therein. The proposals and ideas regarding the (generalized) second law, to date it was written, have been nicely summarized by Aron Wall \cite{Wall:2009wm, Wall:2011hj}.

Our definition of entropy should also be endowed with a notion of the second law. While our local-boost formulation renders the entropy an intrinsic, geometric property of $\Sigma$ detached from specific vector fields, discussing its \textit{variation} inevitably brings the observer back to the forefront. To formulate a law of non-decreasing entropy, one must record its evolution along the trajectory of a physical observer who is equipped with a chosen clock. The thermodynamic description of a gravitating system is profoundly sensitive to the observer's kinematic state, a concept elegantly demonstrated by the Unruh effect \cite{Unruh:1976db}. Although we deal with an observer-independent definition of entropy, the inherently observer-dependent thermodynamic evolution of it leads to a fundamental question: \emph{who writes the second law?} In other words, which class of causal observers, and with which path parametrization, guarantees a universal, non-decreasing gravitational entropy?

This work is a companion to \cite{To-appear-2ndlaw-observer}, which addresses the above question by identifying free-fall observers, with their trajectories described in the ``expansion-adjusted parametrization,'' as the class for which the second law takes a universal form. Here, we provide a more detailed derivation, analyses, and discussions and establish that for these generic causal observers, the second law is fulfilled if the matter content satisfies the integrated (along the observer's path) strong energy condition (SEC). The rest of this paper is organized as follows. Sec.~\ref{sec:geometry} develops the geometric setup; Sec.~\ref{sec:entropy_charge} derives the local-boost entropy charge which yields the area law for Einstein gravity; Sec.~\ref{sec:second_law_derivation} establishes the second law for the relevant causal observers and parametrization, including the null observer limit; and Sec.~\ref{sec:discussion} discusses interpretation and open directions. In Appendix~\ref{apdx:overview} we review the gravitational second law literature.

\vspace*{-5mm}
\section{Geometric Setup and 2+(D-2) Decomposition}\label{sec:geometry}
\vspace*{-2mm}

We begin by considering a generic $D$-dimensional spacetime foliated by codimension-1 slices, one of which is the boundary $\Gamma$, introduced below, alongside its codimension-2 boundary $\Sigma$. In turn, $\Gamma$ admits codimension-2 foliations by $\Sigma$'s. 
In the framework adopted here, the gravitational entropy associated with a spatial region (such as a constant-time slice) is expressed as an integral over its boundary $\Sigma$.

{The local geometry near $\Sigma$ admits a natural decomposition. At each point on the surface, the tangent space splits into the $(D-2)$-dimensional tangent space of $\Sigma$ and a 2-dimensional ($2d$) normal plane, denoted as $\mathcal{D}$. We suppose this entropy is probed by an observer whose worldline is a future-directed causal curve $\gamma$—either timelike or null—emanating orthogonally from $\Sigma$. Since $\gamma$ is everywhere normal to $\Sigma$, the observer's motion is strictly confined to $\mathcal{D}$; hence, the spacetime topology locally decomposes as $\mathcal{D} \times \Sigma$). This geometric setup is illustrated in Fig.~\ref{fig:setup-1}.}

{As the observer evolves along $\gamma$, the family of surfaces  $\Sigma$ sweep out a codimension-1 causal hypersurface $\Gamma$. Locally, $\Gamma$ has the topology $\Sigma \times \gamma$ (see Fig.~\ref{fig:setup-2}) and effectively acts as the causal boundary for the spacetime region under consideration.} {Depending on the specific analysis, it can be more convenient to span the normal plane $\mathcal{D}$ using a pair of null vector fields or an orthogonal basis comprising a timelike and a spacelike vector field. The two choices provide complementary descriptions of the causal geometry in the derivations that follow.}

\vspace*{-6mm}
\subsection{The null decomposition}\label{sec:null_decomposition}
\vspace*{-3mm}

We can span the normal plane $\mathcal{D}$ using two future-oriented null 1-forms (covectors) $l_\mu$ and $n_\mu$, as depicted in Fig.~\ref{fig:setup-1}:
\begin{equation}\label{l-n-frame}
   l\cdot l=0\,,\qquad n\cdot n=0\,,\qquad l\cdot n=-1\,.
\end{equation}
These inner products are defined with respect to the $D$-dimensional spacetime metric $g_{\mu\nu}$. Denoting the induced metric on $\Sigma$ by $q_{\mu\nu}$, we have the following completeness and orthogonality relations:
\begin{equation}\label{g-q-transerve}
   g_{\mu\nu}= q_{\mu\nu}- l_\mu n_\nu - n_\mu l_\nu\,, \qquad  q_{\mu\nu} l^\nu=0=q_{\mu\nu} n^\nu\,,
\end{equation}
where indices are raised or lowered using the full metric $g_{\mu\nu}$. The binormal 2-form to $\Sigma$ is then naturally defined as:
\begin{equation}\label{binormal}
    {\beps}_{\mu\nu}:= l_\mu n_\nu-l_\nu n_\mu\,,\qquad  {\beps}_{\mu\nu} {\beps}^{\mu\nu} =-2\,.
\end{equation}

While the relations above define the local frame $l_\mu, n_\mu$ at a given point, their covariant derivatives (with respect to $g_{\mu\nu}$) do not necessarily remain confined to $\mathcal{D}$. In general, they admit the following decomposition:
\begin{subequations}\label{derv-l-n-general}\begin{align}
 \hspace*{-2mm}    \nabla_{\mu}l_{\nu} &= \Theta^{l}_{\mu\nu} - \kappa_{l}\, n_{\mu}l_{\nu} - \kappa_{n}\, l_{\mu}l_{\nu} +  \mathcal{H}_{\mu} l_{\nu} + l_{\mu}\Omega^{l}_{\nu} + n_{\mu} a_{\nu}^{l}\,, \label{CD-l-inde-general}\\
     \hspace*{-2mm}  \nabla_{\mu}n_{\nu} &=  \Theta^{n}_{\mu\nu} + \kappa_{n}\, l_{\mu}n_{\nu} + \kappa_{l}\, n_{\mu}n_{\nu} - \mathcal{H}_{\mu} n_{\nu} + n_{\mu}\Omega^{n}_{\nu} + l_{\mu} a^{n}_\nu\,. \label{CD-n-inde-general}
\end{align}\end{subequations}
Here, $\Theta^{l,n}_{\mu\nu}$, $\kappa_{l,n}$, and $\mathcal{H}_{\mu}$ denote the deviation tensors (extrinsic curvatures), the non-affinity parameters, and the Hajicek 1-form for the null vectors, respectively (see, e.g., \cite{Gourgoulhon:2005ng}). Additionally, the transverse 1-forms $\Omega^{l,n}_{\nu}$ and $a^{l,n}_{\nu}$ represent the normal fundamental forms (normal connections) and the transverse acceleration vectors.

\textbf{Frame transport and gauge fixing.} We have the freedom to choose how this local frame is transported away from $\Sigma$. This freedom is partially fixed by imposing the following convenient conditions:
\begin{subequations}\label{hypersurface-orthogonality}
\begin{align}
    l_{[\mu} \nabla_\nu l_{\alpha]} \mathrel{\hat{=}} 0\,,&\qquad n_{[\mu} \nabla_\nu n_{\alpha]} \mathrel{\hat{=}} 0\,,\label{hyper-surface-ortho-1}\\ 
    q^{\mu\nu} l^\alpha \nabla_\alpha n_\nu \mathrel{\hat{=}} 0\,,&\qquad  q^{\mu\nu} n^\alpha \nabla_\alpha l_\nu \mathrel{\hat{=}} 0\,,\label{hyper-surface-ortho-2}
\end{align} 
\end{subequations}
where the symbol $\mathrel{\hat{=}}$ indicates that these conditions are only required to hold over the causal boundary $\Gamma$, rather than globally throughout the entire spacetime. 

The hypersurface orthogonality conditions \eqref{hyper-surface-ortho-1} imply that the transverse accelerations vanish ($a_\mu^{l,n} \mathrel{\hat{=}} 0$) and that the deviation tensors $\Theta^{l,n}_{\mu\nu}$ are twist-free (i.e., they have no antisymmetric parts). Consequently, they can be decomposed entirely into a trace part (expansion $\theta$) and a symmetric-traceless part (shear $N_{\mu\nu}$):
\begin{equation}\label{deviation-tensors-decomposed}
     \Theta^{l}_{\mu\nu} \mathrel{\hat{=}} \frac{\theta_{l}}{D-2}q_{\mu\nu}+N^{l}_{\mu\nu}\,,  \qquad \Theta^{n}_{\mu\nu} \mathrel{\hat{=}} \frac{\theta_{n}}{D-2}q_{\mu\nu}+N^{n}_{\mu\nu} \,.
\end{equation}
Meanwhile, the transverse conditions \eqref{hyper-surface-ortho-2} ensure that the null vectors are not rotated outside the normal plane, yielding $\Omega_\mu^{l,n} \mathrel{\hat{=}} 0$. In other words, their parallel transport along one another remains strictly within $\mathcal{D}$. Upon \eqref{hypersurface-orthogonality}, the covariant derivatives evaluated over $\Gamma$ simplify to:
\begin{subequations}\label{derv-l-n-gauge0fixed}\begin{align}
     \nabla_{\mu}l_{\nu} &\mathrel{\hat{=}} \Theta^{l}_{\mu\nu} - \kappa_{l}\, n_{\mu}l_{\nu} - \kappa_{n}\, l_{\mu}l_{\nu} +  \mathcal{H}_{\mu} l_{\nu} \,,\label{CD-l-inde}\\
      \nabla_{\mu}n_{\nu} &\mathrel{\hat{=}}  \Theta^{n}_{\mu\nu} + \kappa_{n}\, l_{\mu}n_{\nu} + \kappa_{l}\, n_{\mu}n_{\nu} - \mathcal{H}_{\mu} n_{\nu}\,. \label{CD-n-inde}
\end{align}\end{subequations}
We emphasize that the transverse 1-forms $a^{l,n}_\mu$ and $\Omega^{l,n}_\mu$ mathematically represent pure gauge degrees of freedom, which can be systematically eliminated through the gauge-fixing conditions applied above. Physically, this is justified because these terms do not enter the null symplectic potential as characteristic initial data \cite{Hopfmuller:2016scf, Chandrasekaran:2018aop, Adami:2021kvx, Adami:2023wbe}.


\begin{figure}[t]
\centering
\begin{tikzpicture}[>=stealth, scale=1.7]
  \fill[gray!10] (0,0) -- (1.4,1.4) -- (-1.4,1.4) -- cycle;
  \fill[gray!10] (0,0) -- (1.4,-1.4) -- (-1.4,-1.4) -- cycle;

  \draw[dashed, gray!80] (-1.4,-1.4) -- (1.4,1.4);
  \draw[dashed, gray!80] (-1.4,1.4) -- (1.4,-1.4);

  \draw[->, thick, gray!70!black] (-1.7,0) -- (1.7,0) node[right, text=black] {$r$};
  \draw[->, thick, gray!70!black] (0,-1.7) -- (0,1.7) node[above, text=black] {$\tau$};

  \draw[very thick, blue!80!black, decoration={markings, mark=at position 0.85 with {\arrow{>}}}, postaction={decorate}]
    plot[domain=-1.4:1.4, samples=100, variable=\t] 
    ({0.5*\t - 0.2917*ln(0.5*(exp(1.2*\t) + exp(-1.2*\t)))}, {\t})
    node[right] {\large $\gamma$};

  \draw[black!60, thick] (-0.13, 0.07) -- (-0.06, 0.20) -- (0.07, 0.13);

  \draw[->, very thick, red!80!black] (0,0) -- (0.8,0.8) node[right] {$l^\mu$};
  \draw[->, very thick, red!80!black] (0,0) -- (-0.8,0.8) node[left] {$n^\mu$};

  \draw[->, very thick, cyan!80!black] (0,0) -- (0.45, 0.90)  node[above right=-2pt, inner sep=1pt, fill=white, fill opacity=0.7, text opacity=1] {$v_{\lambda}^{\mu}$};
  \draw[->, very thick, green!60!black] (0,0) -- (-0.90, 0.45) node[left] {$s^\mu$};

  \filldraw[black] (0,0) circle (1.5pt);
  \node[below right=2pt, font=\bfseries] at (0,0) {$\Sigma(\lambda)$};

  \begin{scope}[shift={(1.3, 1.4)}]
    \node at (0.2, 0.3) {$\mathcal{D}$};
    \draw[->, thick] (0,0) -- (0.4,0);
    \draw[->, thick] (0,0) -- (0,0.4);
  \end{scope}
\end{tikzpicture}
\caption{\justifying{The future-oriented causal trajectory of the observer $\gamma$ in the $2d$ plane $\mathcal{D}$. The local frame is adapted such that the tangent vector $v^{\mu}_{\lambda}$ and the spacelike normal vector $s^\mu$ are orthogonal. As $\lambda \to \infty$, the curve asymptotes to a constant radius, representing an equilibrium state with vanishing expansion $\theta_v$.}}
\label{fig:setup-1}
\end{figure}
\vspace*{-7mm}
\subsection{Parameterization of the causal boundary $\Gamma$}\label{sec:causal_boundary}
\vspace*{-3mm}

{Thus far, we have left the coordinate choices on $\Sigma$ and $\mathcal{D}$ unspecified, while only partially fixed the null basis $\{l_\mu, n_\mu\}$. We should still need to more precisely and systematically define the observer path $\gamma$.  To this end, consider a reference observer who follows a causal trajectory parameterized by proper time $\tau$, $x^\mu = x^\mu(\tau)$. In this \textit{reference parametrization}, the future-directed causal velocity field of the observer $v^\mu_\tau$ is locally decomposed in the null basis as (cf. Fig.~\ref{fig:setup-1}):}
\begin{equation}\label{causal-vector}
    v^{\mu}_{\tau} :=\frac{\d{} x^\mu}{\d{}\tau} = \frac{1}{\sqrt{2}}(l^{\mu} + n^{\mu} )\,,\qquad 
    v^{2}_{\tau}= -1\,.
\end{equation}

\begin{figure}[H]
\centering
\begin{tikzpicture}[>=stealth, scale=1.15]

  \begin{scope}[shift={(0,0)}]
    \fill[gray!10] (0,0) -- (1.8,1.8) -- (-1.8,1.8) -- cycle;
    
    \draw[dashed, gray!80] (0,0) -- (1.8,1.8);
    \draw[dashed, gray!80] (0,0) -- (-1.8,1.8);
    
    \draw[->, very thick, red!80!black] (0,0) -- (1.4,1.4) node[right] {$l^\mu$};
    \draw[->, very thick, red!80!black] (0,0) -- (-1.4,1.4) node[left] {$n^\mu$};
    
    \draw[dashed, black!40, thick] plot[domain=-1.0:1.0, samples=50] ({sinh(\x)}, {cosh(\x)});
    \node[cyan!80!black] at (-1.3, 0.9) {};

    \draw[->, very thick, gray] (0,0) -- (0,1) node[below right, text=black] {$v_\tau$};
    \filldraw[gray] (0,1) circle (1.2pt);
    
    \draw[->, very thick, cyan!80!black] (0,0) -- (0.758, 1.255) node[above left=-2pt] {$v$};
    \filldraw[cyan!80!black] (0.758, 1.255) circle (1.2pt);
    
    \draw[->, thick, palatinateblue] (0.12, 1.05) to[bend right=25] node[midway, above left=-0.5pt, inner sep=1pt] {\footnotesize $\mathscr{B}$} (0.68, 1.22);
    
    \node[align=center, font=\footnotesize, text=black!80] at (0, -0.7) {
        \textbf{(a) Geometric Steering} \\ 
        $\mathscr{B}$ tilts the vector along the hyperbola \\
        to yield an affinely parameterized \\
        geodesic ($\kappa_v \mathrel{\hat{=}} 0$).
    };
  \end{scope}

  \begin{scope}[shift={(0,-4.4)}]
    \fill[gray!10] (0,0) -- (1.8,1.8) -- (-1.8,1.8) -- cycle;
    
    \draw[dashed, gray!80] (0,0) -- (1.8,1.8);
    \draw[dashed, gray!80] (0,0) -- (-1.8,1.8);
    
    \draw[->, very thick, red!80!black] (0,0) -- (1.4,1.4) node[right] {$l^\mu$};
    \draw[->, very thick, red!80!black] (0,0) -- (-1.4,1.4) node[left] {$n^\mu$};
    
    \draw[dashed, black!40, thick] plot[domain=-1.0:1.0, samples=50] ({sinh(\x)}, {cosh(\x)});
    
    \draw[dashed, amaranth, thick] plot[domain=-0.8:0.8, samples=50] ({1.5*sinh(\x)}, {1.5*cosh(\x)});
    \node[amaranth,blue!80!black] at (-0.55, 1.9) {\scriptsize $v_{\lambda}^2 = -\mathscr{R}^2$};
    
    \draw[->, very thick, blue!80!black] (0,0) -- (1.137, 1.882) node[above left=-2pt] {$v_\lambda$};
    \filldraw[blue!80!black] (1.137, 1.882) circle (1.2pt);

    \draw[->, very thick, cyan!80!black] (0,0) -- (0.758, 1.255);
    \filldraw[cyan!80!black] (0.758, 1.255) circle (1.2pt) node[right, text=cyan!80!black] {$v$};
    
    \draw[->, thick, amaranth] (0.83, 1.35) -- (1.05, 1.75) node[midway, right] {\footnotesize $\mathscr{R}$};
    
    \node[align=center, font=\footnotesize, text=black!80] at (0, -0.7) {
        \textbf{(b) Clock Calibration} \\ 
        $\mathscr{R}$ scales the tangent vector \\ 
        to conserve the boundary proper \\
        volume ($v \cdot \nabla \ln\big(\sqrt{q}\mathscr{R}\big) \mathrel{\hat{=}} 0$).
    };
  \end{scope}

\end{tikzpicture}
\caption{\justifying A depiction of the physical and geometric meanings of the kinematic freedoms $\mathscr{B}$ and $\mathscr{R}$ in the transverse plane $\mathcal{D}$. \textbf{(a)} The local boost parameter $\mathscr{B}$ glides the generic future-directed causal vector $v_\tau$ along the unit hyperbola $v^2=-1$. The specific choice of $\mathscr{B}$ in \eqref{ansatz-B-R-2} aligns the normal frame such that the observer's trajectory is an affinely parameterized geodesic. \textbf{(b)} The reparameterization function $\mathscr{R}$ scales the length of the vector, shifting it to a new hyperbola $v_\lambda^2 = -\mathscr{R}^2$. The choice of $\mathscr{R}$ in \eqref{ansatz-B-R} precisely balances the clock ticks against the spatial expansion $\theta_v$, ensuring that the codimension-1 volume form $\sqrt{q}\mathscr{R}$ is conserved along the causal boundary $\Gamma$.}
\label{fig:boost_reparam}

\makeatletter
\protected@edef\@currentlabel{\thefigure(a)}\label{fig:boost_reparam_a}
\protected@edef\@currentlabel{\thefigure(b)}\label{fig:boost_reparam_b}
\makeatother

\end{figure}



The most general causal observer can be constructed from this reference vector $v_\tau^\mu$ by exploiting two local freedoms:
\begin{enumerate}[leftmargin=4mm]
    \item[(1)] \textbf{Reparameterization} of the path, $\tau\to \lambda$, under which the tangent vector scales as (see Fig.~\ref{fig:boost_reparam_a}):
    \begin{equation}\label{reparametrization}
        l^{\mu}\to  \mathscr{R}\,  l^{\mu}\,, \qquad  n^{\mu}\to \mathscr{R}\, n^{\mu}\,, \qquad \mathscr{R} := \frac{\text{d} \tau}{\text{d} \lambda}\,.
    \end{equation}
We require $\mathscr{R}>0$ to ensure that the reparameterization preserves the future-oriented nature of the causal curve. {Physically, the choice of $\mathscr{R}$ corresponds to a local calibration of the observer's clock along their trajectory.}

    \item[(2)] \textbf{Local boosts} in the  plane $\mathcal{D}$ (see Fig.~\ref{fig:boost_reparam_b}). The defining relations \eqref{l-n-frame} and \eqref{hypersurface-orthogonality} fix the frame only up to local Lorentz boosts, under which:
    \begin{equation}\label{local-boost}
        l^\mu\to \mathscr{B} l^\mu\,,\qquad n^\mu\to \mathscr{B}^{-1} n^\mu\,, \qquad \mathscr{B} >0\,.
    \end{equation}
\end{enumerate}

Applying both the path reparameterization \eqref{reparametrization} and the local boost \eqref{local-boost} to \eqref{causal-vector}, we obtain
\begin{equation}\label{causal-vector-RB}
    v^\mu_{\lambda} := \frac{\d{}x^\mu}{\d{}\lambda} = \mathscr{R}\, v^\mu\, ,\qquad v^\mu = \frac{1}{\sqrt{2}} (\mathscr{B} l^\mu +\mathscr{B}^{-1} n^\mu)\, ,
\end{equation}
where $v^2 = - 1, \ v_\lambda^2 = - \mathscr{R}^2$ and  $v^\mu_{\lambda}$ is the tangent vector along the physical causal curve $\gamma$, now parameterized by $\lambda$ (see Fig.~\ref{fig:setup-1}). By construction, $v^{\mu}_{\lambda}$ is tangent to the codimension-1 causal hypersurface $\Gamma$ (see Fig.~\ref{fig:setup-2}). At any fixed $\lambda$, $\Gamma$ is bounded by the spacelike codimension-2 surface $\Sigma(\lambda)$.

The corresponding spacelike unit normal to $\Gamma$, which lies entirely within  $\mathcal{D}$, is given by:
\begin{equation}\label{vec-s}
    s^\mu := \frac{1}{\sqrt{2}}(-\mathscr{B} l^\mu +\mathscr{B}^{-1} n^\mu)\, , \qquad s^2=+1\,, 
\end{equation}
which readily satisfies $v_{\lambda} \cdot s=0$. We note that $\Gamma$ is a timelike (or null, in the strict null limit) codimension-1 surface. In contrast, the hypersurface generated by parallel transporting $\Sigma(\lambda)$ along the normal vector field $s^\mu$ forms a partial Cauchy surface $\mathcal{C}$. In our subsequent analysis, we will explicitly integrate physical fluxes over the causal history boundary $\Gamma$ rather than the spacelike slice $\mathcal{C}$. We reiterate that our gauge-fixed relations \eqref{derv-l-n-gauge0fixed} need only hold over $\Gamma$, not globally throughout the $D$-dimensional spacetime.


\begin{figure}[H]
\centering
\begin{tikzpicture}[scale=.8, >=stealth,
    surface/.style={left color=gray!30, right color=gray!15, middle color=white, opacity=0.6},
    boundary/.style={thick, draw=black!75},
    hidden/.style={dashed, draw=black!50, thick},
    slicefill/.style={left color=blue!30, right color=blue!10, opacity=0.5}
  ]

  \def\H{4.5}                  
  \def\Rb{1.3} \def\ryb{0.35}  
  \def\Rt{2.4} \def\ryt{0.6}   

  \def\Rm{2.04}   
  \def\rym{0.51} 
  \def\Ym{2.25}   

  \shade[surface]
    (-\Rb, 0) .. controls (-1.8, 1.5) and (-2.4, 3.0) .. (-\Rt, \H) 
    arc [start angle=180, end angle=360, x radius=\Rt, y radius=\ryt]
    .. controls (2.4, 3.0) and (1.8, 1.5) .. (\Rb, 0) 
    arc [start angle=360, end angle=180, x radius=\Rb, y radius=\ryb];

  \draw[hidden] (\Rb, 0) arc [start angle=0, end angle=180, x radius=\Rb, y radius=\ryb];
  \draw[hidden, draw=blue!50!black] (\Rm, \Ym) arc [start angle=0, end angle=180, x radius=\Rm, y radius=\rym];
  \draw[hidden] (\Rt, \H) arc [start angle=0, end angle=180, x radius=\Rt, y radius=\ryt];

  \shade[slicefill] (0, \Ym) ellipse [x radius=\Rm, y radius=\rym];

  \draw[boundary] (-\Rb, 0) .. controls (-1.8, 1.5) and (-2.4, 3.0) .. (-\Rt, \H);
  \draw[boundary] (\Rb, 0) .. controls (1.8, 1.5) and (2.4, 3.0) .. (\Rt, \H);

  \draw[boundary] (-\Rb, 0) arc [start angle=180, end angle=360, x radius=\Rb, y radius=\ryb];
  \draw[boundary, draw=blue!60!black] (-\Rm, \Ym) arc [start angle=180, end angle=360, x radius=\Rm, y radius=\rym];
  \draw[boundary] (-\Rt, \H) arc [start angle=180, end angle=360, x radius=\Rt, y radius=\ryt];

  
  \node[font=\Large\bfseries, text=black!80] at (-2.6, 3.5) {$\Gamma$};

  \coordinate (SigmaPointer) at ({-\Rm + 0.15}, {\Ym - 0.1});
  \draw[<-, thick, blue!60!black] (SigmaPointer) to[out=210, in=0] ++(-0.8, -0.4) node[left] {\large $\Sigma(\lambda)$};

  \draw[very thick, blue!80!black, decoration={markings, mark=at position 0.3 with {\arrow[scale=1.2]{>}}}, postaction={decorate}]
    plot[domain=0:1, samples=40, variable=\t]
    ( { (1.3*(1-\t)*(1-\t)*(1-\t) + 5.4*(1-\t)*(1-\t)*\t + 7.2*(1-\t)*\t*\t + 2.4*\t*\t*\t)*0.866 },
      { 4.375*\t - 0.175 }
    );

  \node[blue!80!black] at (1.8, 0.7) {\large $\gamma$};

  \coordinate (Gmid) at (1.764, 2.012);
  \fill[black] (Gmid) circle (1.5pt);

  \def\vx{0.28} \def\vy{1.09} 
  \def\sx{0.87} \def\sy{-0.17} 

  \pgfmathsetmacro{\mkvx}{0.18 * \vx}
  \pgfmathsetmacro{\mkvy}{0.18 * \vy}
  \pgfmathsetmacro{\mksx}{0.18 * \sx}
  \pgfmathsetmacro{\mksy}{0.18 * \sy}
  
  \draw[thick, black!60, line cap=round, line join=round]
    ($(Gmid) + (\mkvx, \mkvy)$) --
    ($(Gmid) + (\mkvx + \mksx, \mkvy + \mksy)$) --
    ($(Gmid) + (\mksx, \mksy)$);

  \draw[->, very thick, cyan!80!black] (Gmid) -- ++(\vx, \vy) node[right=2pt, font=\bfseries] {$v^{\mu}$};

  \draw[->, very thick, green!60!black] (Gmid) -- ++(\sx, \sy) node[below right=-1pt, font=\bfseries] {$s^\mu$};

\end{tikzpicture}
\caption{The codimension-1 causal surface $\Gamma$ which is topologically $\Sigma \times \gamma$. At a given $\lambda$, $\Gamma$ is limited to the spacelike boundary $\Sigma(\lambda)$. As $\lambda \to \infty$, the surface reaches an equilibrium state where the expansion vanishes ($\theta_v \to 0$).}
\label{fig:setup-2}
\end{figure}


\textbf{Behavior under local boosts.} While the frame definitions \eqref{l-n-frame} and \eqref{hypersurface-orthogonality} are explicitly invariant under the local boosts \eqref{local-boost}, the geometric coefficients on the right-hand side of \eqref{derv-l-n-gauge0fixed} are not. They transform as:
\begin{subequations}\label{gamma-scaling}
\begin{align}
   \theta_l \to  \mathscr{B} \theta_l, &\quad \quad \theta_n \to \mathscr{B}^{-1} \theta_n,\\    
 {N^{l}_{\mu\nu} \to \mathscr{B} N^{l}_{\mu\nu}}, &\quad\quad {N^{n}_{\mu\nu} \to \mathscr{B}^{-1} N^{n}_{\mu\nu}} \,,\\
 \hspace*{-3mm}  \kappa_{l} \to \mathscr{B}( \kappa_{l} + l\cdot\nabla \ln\mathscr{B}), &\, \, \,  \kappa_{n} \to \mathscr{B}^{-1}( \kappa_{n} + n\cdot \nabla \ln\mathscr{B}).
 \end{align}
\end{subequations}
This implies the following useful relation:
\begin{equation}\label{kappa-ln}
    \kappa_{l} n_\mu + \kappa_{n} l_\mu \to \kappa_{l} n_\mu + \kappa_{n} l_\mu - \nabla_\mu \ln\mathscr{B}\,.
\end{equation}
Finally, for completeness, we note that the Hajicek 1-form transforms as $\mathcal{H}_{\mu} \to \mathcal{H}_{\mu} + q_\mu{}^\nu \nabla_\nu \ln\mathscr{B}$. If one imposes $q_\mu{}^\nu \nabla_\nu \ln\mathscr{B}=0$—which can be achieved by fixing the remaining rotational freedoms on $\Sigma$—then $\mathcal{H}_{\mu}$ becomes strictly boost invariant. 

In what follows, we will leverage \eqref{kappa-ln} and this local boost freedom to enforce specific physical conditions along the observer's trajectory.

\vspace*{-6mm}
\subsection{The timelike and spacelike decomposition}\label{sec:timelike_decomposition}
\vspace*{-3mm}

Alternatively, we can span $\mathcal{D}$ using the timelike unit vector $v_\mu$ \eqref{causal-vector-RB} and the spacelike unit vector $s_\mu$ \eqref{vec-s}. In this basis, the spacetime metric decomposes as:
\begin{equation}\label{metric-v-s-decomp}
    g_{\mu\nu}=-v_{\mu}v_{\nu}+s_{\mu}s_{\nu}+q_{\mu\nu}\, .
\end{equation}
Translated into the $(v, s)$ basis, the hypersurface orthogonality and in-plane derivative conditions \eqref{hypersurface-orthogonality} take the form:
\begin{equation}\label{hyper-ortho-v-s}
 \begin{split}
     & a_{\alpha}^{v} := v^{\mu} q^{\nu}_{\alpha} \nabla_{\mu}v_{\nu} \mathrel{\hat{=}} 0\, , \quad \eta^{v}_\alpha :=- s^{\mu}  q^{\nu}_{\alpha} \nabla_{\mu}v_{\nu} \mathrel{\hat{=}} 0\, , \\
     & a_{\alpha}^{s} :=  s^{\mu}  q^{\nu}_{\alpha} \nabla_{\mu}s_{\nu} \mathrel{\hat{=}} 0 \, , \quad \eta^{s}_\alpha :=- v^{\mu} q^{\nu}_{\alpha} \nabla_{\mu} s_{\nu} \mathrel{\hat{=}} 0\, .
 \end{split}
\end{equation}
Imposing these conditions, the covariant derivatives of the basis vectors evaluated over $\Gamma$ simplify significantly:
\begin{subequations}
   \begin{align}
    & \nabla_\mu v_\nu \mathrel{\hat{=}} \kappa_v v_\mu s_\nu - \kappa_s s_\mu s_\nu - \omega_\mu s_\nu + \Theta_{\mu\nu}^{(v)} \,, \label{cov-v-decompo} \\
    & \nabla_\mu s_\nu \mathrel{\hat{=}}  \kappa_v v_\mu v_\nu -\kappa_s s_\mu v_\nu  - v_\nu \omega_\mu + \Theta^{(s)}_{\mu\nu}\, . \label{cov-s-decompo}
   \end{align}
\end{subequations}
Here, $\omega_\mu$ is a strictly transverse 1-form ($v^\mu\omega_\mu=0=s^\mu\omega_\mu$). The geometric variables associated with $v$ and $s$ are linear combinations of their null counterparts, explicitly coupled to the local boost parameter $\mathscr{B}$:
\begin{subequations}
    \begin{align}
       & \Theta_{\mu\nu}^{v} = \frac{1}{\sqrt{2}}\big(\mathscr{B} \Theta_{\mu\nu}^l+\mathscr{B}^{-1} \Theta_{\mu\nu}^n\big) = \frac{\theta_v}{D-2} q_{\mu\nu}+ N_{\mu\nu}^v\, , \label{devi-tensor-v}\\
       & \Theta_{\mu\nu}^{s} = \frac{1}{\sqrt{2}} \left(\mathscr{B}^{-1} \Theta^{n}_{\mu\nu} - \mathscr{B} \Theta^{l}_{\mu\nu} \right) =  \frac{\theta_s}{D-2} q_{\mu\nu}+ N_{\mu\nu}^s\, ,\label{devi-tensor-s} \\
       & \kappa_v = \frac{1}{\sqrt{2}}\left(\mathscr{B}\kappa_{l}+\mathscr{B}^{-1}\kappa_{n}\right) + v \cdot\nabla\ln\mathscr{B}\, , \label{kappa-v} \\
       & \kappa_s = \frac{1}{\sqrt{2}} \left( -\mathscr{B}\, \kappa_l + \mathscr{B}^{-1} \kappa_n\right)  + s\cdot\nabla \ln \mathscr{B}\, , \label{kappa-s}\\
       & \omega_\mu = \mathcal{H}_\mu + q_{\mu \nu} \nabla^{\nu} \ln \mathscr{B}\, .\label{rotation-1form} 
    \end{align}
\end{subequations}
Crucially, reading from \eqref{cov-v-decompo}, we observe that if $\kappa_v \mathrel{\hat{=}} 0$, the vector $v^\mu$ becomes the velocity vector of an affinely parameterized geodesic, satisfying $v^\mu\nabla_\mu v^\nu \mathrel{\hat{=}} 0$.

\vspace*{-4mm}
\section{Gravitational Entropy as a Lorentz Boost Charge}\label{sec:entropy_charge}

{Having established the geometric setup, we now formalize our definition of gravitational entropy. Following the seminal work of Wald \cite{Wald:1993nt, Iyer:1994ys}, the standard literature typically defines entropy in generic diffeomorphism-invariant theories of gravity as a charge associated with a null vector symmetry generator (a diffeomorphism) evaluated on a null boundary. In this conventional viewpoint, entropy is fundamentally treated as an energy divided by a temperature (or surface gravity). This is made explicit in Wald's entropy formula for Killing horizons, where the entropy is defined as the Noether charge associated with the horizon-generating Killing vector field, normalized by the surface gravity \cite{Wald:1993nt}. (As a zeroth-order consistency check, this construction successfully recovers the Bekenstein-Hawking area law for Einstein gravity.)}

{However, invoking the covariant phase space formalism (CPSF) reveals that this division by surface gravity can introduce integrability issues \cite{Hajian:2015xlp}. Conceptually, it also perpetually ties the definition of entropy to the notion of energy. These issues persist well beyond static Killing horizons, appearing in contexts with dynamical horizons or generic dynamical entropy \cite{Ashtekar:2002ag, Booth:2006bn, Wall:2015raa, Hollands:2022fkn, Hollands:2024vbe, Visser:2024pwz, Shajiee:2026coz}. Fundamentally, it is highly desirable to disentangle a concept as foundational as entropy from energy at the very level of its definition.}

{To bypass these issues, we note that they primarily stem from utilizing a spacetime vector field as the symmetry generator to compute a charge over a codimension-2 surface. In this work, we adopt a different approach within the CPSF: we define the gravitational entropy of a spatial region enclosed by a codimension-2 surface $\Sigma$ as the surface charge associated strictly with \textit{local Lorentz boosts} in the transverse $2d$ plane $\mathcal{D}$.  This viewpoint relies on an internal gauge transformation rather than a spacetime vector field. Thus, it evades the ``normalization issue,'' and as we show explicitly below, it directly yields a robust, integrable entropy.}

\vspace*{-6mm}
\subsection{Review of Covariant Phase Space Formalism (CPSF)}\label{sec:cpsf_review}
\vspace*{-3mm}

The CPSF provides a mathematically robust framework for defining and computing conserved charges associated with continuous symmetries, making it particularly useful for diffeomorphism-invariant theories \cite{Lee:1990nz, Barnich:2007bf, Harlow:2014yka, Compere:2018aar, Grumiller:2022qhx, Golshani:2026lvc}. Here, we apply it to gravity theories in the first-order formulation \cite{MacDowell:1977jt, Hehl:1994ue}, which, besides diffeomorphism invariance, exhibits manifest invariance under local Lorentz transformations.

Here we adopt the notation used in \cite{Wald:1993nt, Iyer:1994ys}. Consider an action $\mathcal{I}$ constructed from a Lagrangian density $D$-form $\mathbf{L}$ on a spacetime manifold $\mathcal{M}$. An arbitrary first-order variation of the Lagrangian yields the equations of motion $\mathbf{E}_i$ alongside a total-derivative boundary symplectic potential $(D-1)$-form $\boldsymbol{\Theta}$:
\begin{equation}\label{var-Lag-L}
  \delta\mathbf{L} = \mathbf{E}_{i}\delta\Phi^{i} + \d{}\boldsymbol{\Theta}[\Phi,\delta\Phi]\,,
\end{equation}
where $\Phi^{i}$ denotes the collection of dynamical fields.  

To extract charges, we consider a generic symmetry generator $\zeta$, which represents a transformation $\Phi\to \Phi+\delta_\zeta\Phi$ under which the Lagrangian is invariant up to a total derivative. In the first-order formulation of gravity, $\zeta$ encompasses both spacetime diffeomorphisms $\xi$ and internal gauge transformations (e.g., local Lorentz transformations $\Lambda$), such that $\zeta=\{\xi,\Lambda\}$. Under such a transformation, the Lagrangian varies purely via the Lie derivative:
\begin{equation}
    \delta_\zeta \mathbf{L} = \mathcal{L}_\xi \mathbf{L} = \d{}(\iota_\xi \mathbf{L})\, ,
\end{equation} 
where $\iota_\xi$ is the interior product along the vector field $\xi$.

Equating this to \eqref{var-Lag-L} evaluated for $\delta = \delta_\zeta$, we find that on-shell ($\mathbf{E}_i = 0$), the Noether current $(D-1)$-form is conserved:
\begin{equation}\label{Noether-current}
    \mathbf{J}[\zeta] := \boldsymbol{\Theta}[\delta_\zeta] - \iota_\xi \mathbf{L}\,, \qquad \d{}\mathbf{J}[\zeta] \stackrel{\circ}{=} 0\,,
\end{equation}
where $\stackrel{\circ}{=}$ denotes on-shell equality. By the Poincaré lemma, this on-shell conserved current can be written locally as the exact exterior derivative of a Noether charge $(D-2)$-form $\mathbf{Q}_{\rm N}[\zeta]$:
\begin{equation}\label{Noether-charge-density}
\mathbf{J}[\zeta] \stackrel{\circ}{=}  \d{}\mathbf{Q}_{\rm N}[\zeta]\, .   
\end{equation}

The phase space is further equipped with a pre-symplectic current, defined by taking an anti-symmetrized second variation of the symplectic potential: 
\begin{equation}
\boldsymbol{\omega}[\delta_1, \delta_2] = \delta_1 \boldsymbol{\Theta} [\delta_2] - \delta_2 \boldsymbol{\Theta}[\delta_1]\, .   
\end{equation}
When one variation is a symmetry transformation $\delta_\zeta$ and the other is an arbitrary generic variation $\delta$, the fundamental identity of the CPSF (utilizing Cartan's magic formula $\mathcal{L}_\xi \boldsymbol{\Theta}= \d{}(\iota_\xi \boldsymbol{\Theta}) + \iota_\xi \d{}\boldsymbol{\Theta}$) allows us to express the pre-symplectic current as a total derivative:
\begin{equation}
    \boldsymbol{\omega}[\delta, \delta_\zeta] \stackrel{\circ}{=}\d{} \big( \delta Q_{\rm N}[\zeta] - \iota_\xi \boldsymbol{\Theta}[\delta] \big)\,.
\end{equation}
Integrating this over a codimension-1 surface—such as our causal boundary $\Gamma$—localizes the variation of the surface charge to its codimension-2 boundary $\partial \Gamma=\Sigma$:
\begin{equation}\label{usual-charge-var}
    \slashed{\delta} {Q}_\zeta \stackrel{\circ}{=} \int_{\Sigma} \big( \delta \mathbf{Q}_{\rm N}[\zeta] - \iota_{\xi} \boldsymbol{\Theta}[\delta] \big)\,,
\end{equation}
where $\slashed{\delta} Q_\zeta$ denotes a generic 1-form over the solution space. In general, this variation is not exact. The charge is said to be \textit{integrable} if $\slashed{\delta} Q_\zeta$ forms an exact 1-form on phase space. When integrable, there exists a well-defined, finite surface charge $Q_\zeta[\Phi]$ such that $\delta Q_\zeta[\Phi]= \slashed{\delta} Q_\zeta$. (For a detailed and mathematically rigorous discussion, see, e.g., \cite{Golshani:2026lvc}.)

\vspace*{-6mm}
\subsection{Entropy as the charge of local boosts}\label{sec:entropy_local_boosts}
\vspace*{-3mm}

We propose that gravitational entropy is the integrable surface charge associated with local Lorentz boosts in the transverse $2d$ plane $\mathcal{D}$. Our construction relies explicitly on local Lorentz symmetry, making the first-order (metric-affine) formulation of gravity the natural setting. In this framework, the dynamical fields are the frame field 1-forms $e^a$ and the spin connection 1-forms $\omega^{ab}$ (with internal Lorentz indices $a,b=0,1,\cdots, D-1$). For illustrative purposes and to establish the core ideas and results, we focus here on the Einstein-Hilbert theory, though the framework readily applies to generic higher-curvature gravity theories and can be systematically extended to them.

The Einstein-Hilbert action in the first-order formalism is given by the Lagrangian:
\begin{equation}\label{Palatini-D-matter}
    \mathbf{L} = \frac{1}{16\pi G} \star(e^{a}\wedge e^{b})\wedge R_{ab} + 16\pi G\, \star(1) \mathcal{L}_{\text{matter}}\,,
\end{equation}
where $\star$ is the Hodge dual and 
\begin{equation}
R_{ab} = \d{}\omega_{ab} + \omega_{a}{}^{c} \wedge \omega_{cb} = D\omega_{ab}\, ,
\end{equation}
is the curvature 2-form. The first variation of the action identifies the bulk equations of motion 
\begin{align}
    &\star(e_{a} \wedge e_{b} \wedge e_{c})\wedge R^{bc}  + 16 \pi G\, \star(\mathcal{T}_{a})=0 \, ,\label{EoM-e}\\
    &\mathrm{D} \star(e_{a}\wedge e_{b}) = 0\, , \label{EoM-w}
\end{align}
where $\mathrm{D}$ and $\mathcal{T}_{a}$ denote the Lorentz-covariant exterior derivative and the energy-momentum 1-form, respectively. The boundary symplectic potential $(D-1)$-form reads:
\begin{equation}\label{Symplectic-Theta}
    \boldsymbol{\Theta}[\delta] = \frac{1}{16\pi G} \delta\omega^{ab} \wedge \star(e_a \wedge e_b)\,.
\end{equation}

Under a combined symmetry transformation generated by $\zeta=\{\xi,\Lambda\}$, the dynamical fields vary as:
\begin{subequations}\label{var-Xi-e}
\begin{align}
   & \delta_{\zeta} e^a = \mathcal{L}_\xi e^a + \Lambda^a{}_{b}\, e^b \, , \\
   & \delta_{\zeta} \omega^{a}{}_{b} = \mathcal{L}_\xi \omega^{a}{}_{b} - \mathrm{D}\Lambda^{a}{}_{b} \, . \label{omega-trans}
\end{align}
\end{subequations}
To compute the Noether charge $Q_{\rm N}[\zeta]$ using the algorithm from the previous section \ref{sec:cpsf_review}, we systematically rewrite the variation of the connection using Cartan's formula as
\begin{equation}\label{omega-var}
    \delta_\zeta \omega^{ab} = \iota_\xi R^{ab} + \mathrm{D}(\iota_\xi \omega^{ab} - \Lambda^{ab})\,.
\end{equation}
Substituting this into $\boldsymbol{\Theta}[\delta_\zeta]$ and applying the on-shell definitions \eqref{Noether-current} and \eqref{Noether-charge-density}, we obtain the $(D-2)$-form Noether charge density for the first-order Einstein-Hilbert theory:
\begin{equation}\label{Noether-Charge-Palatini}
    \mathbf{Q}_{\rm N}[\zeta] = \frac{1}{16\pi G} (\iota_\xi \omega^{ab} - \Lambda^{ab}) \star(e_a \wedge e_b)\,.
\end{equation}
Inserting this $\mathbf{Q}_{\rm N}[\zeta]$ into the general surface charge variation formula \eqref{usual-charge-var}, we arrive at:
\begin{align}\label{charge-var-general}
    \slashed{\delta} Q_{\zeta} &= \frac{1}{16\pi G} \int_{\Sigma} \Big[ \delta\left((\iota_{\xi}\omega^{ab}-\Lambda^{ab}) \star(e_{a}\wedge e_{b}) \right) \nonumber\\
    &\qquad\qquad\quad - \iota_{\xi}\left(\delta\omega^{ab}\wedge \star(e_{a}\wedge e_{b})\right) \Big] \nonumber\\
    &= \frac{1}{16\pi G} \int_{\Sigma} \Big[ (\iota_{\xi}\omega^{ab}-\Lambda^{ab}) \delta\star(e_{a}\wedge e_{b}) \nonumber\\
    &\qquad\qquad\quad - \delta\omega^{ab}\wedge\iota_{\xi}\star(e_{a}\wedge e_{b}) \Big] \,,
\end{align}
where in the final equality we utilized the field-independence of the symmetry generator ($\delta \zeta = 0$).

To isolate the contribution to the entropy, we restrict our attention strictly to pure local Lorentz transformations by turning off the diffeomorphisms ($\xi=0$, hence $\iota_\xi = 0$) \cite{Godazgar:2020kqd, Jacobson:2015uqa}. Equation \eqref{charge-var-general} then drastically simplifies to:
\begin{align}\label{charge-var-in-D}
    \slashed{\delta} Q_{\Lambda} &= -\frac{1}{16\pi G}\int_{\Sigma}\Lambda^{ab} \delta \big(\star(e_{a}\wedge e_{b})\big) \nonumber\\
    &= -\frac{1}{16\pi G}\delta\left(\int_{\Sigma}\star(e_{a}\wedge e_{b})\,\Lambda^{ab}\right)\,.
\end{align}
Because the symmetry generator $\Lambda^{ab}$ is assumed to be field-independent ($\delta\Lambda^{ab}=0$), the variation can be pulled entirely outside the integral. The charge variation $\slashed{\delta} Q_{\Lambda}$ is therefore \textit{manifestly integrable}. 

We now specify the generator for local boosts in the $2d$ plane $\mathcal{D}$ transverse to $\Sigma$. Identifying the null vectors $l_\mu$ and $n_\mu$ with two of our frame fields ($e^+_\mu = l_\mu$ and $e^-_\mu = n_\mu$), the local boost generator is proportional to the binormal $\beps^{ab}$:
\begin{equation}\label{local-boost-generator}
   \Lambda^{ab}_{\text{boost}}=-\frac{2\pi}{\hbar}\beps^{ab}\,.
\end{equation}
The prefactor $2\pi/\hbar$ is fixed by algebraic requirements. Inserting \eqref{local-boost-generator} into \eqref{charge-var-in-D}, the integrated exact surface charge yields our definition of gravitational entropy:
\begin{equation}\label{charge-in-D-entropy}
  \hspace*{-2mm}S = \frac{1}{4 G} \int_{\Sigma}{\cal S} \, , \quad {\cal S} := \frac{ \varepsilon_{a_1 a_2 a_3 \dots a_D} }{2 \hbar{(D-2)!}} {\beps}^{a_1 a_2} \, e^{a_3} \wedge \dots \wedge e^{a_D}.  
\end{equation}
Consequently, we obtain:
\begin{equation}\label{entropy-formula}
     S = \frac{1}{4 G \hbar} \int_{\Sigma} \d{}^{D-2}x\, \sqrt{q}= \frac{\text{Area}(\Sigma)}{4G\hbar}\, .
\end{equation}
Our construction, not unexpectedly, has given back the standard Bekenstein-Hawking area law for Einstein gravity. Nonetheless, there are important conceptual and technical differences with the standard derivations:
\begin{enumerate}[leftmargin=4mm, itemsep=0mm]
\item[I.] In our derivation, the entropy is an integrable surface charge associated with local Lorentz boosts in the $2d$ plane transverse to \textit{any} codimension-2 surface $\Sigma$; $\Sigma$ need not be a bifurcation surface, and we are not considering any horizon or any null surface.
\item[II.] The symmetry generator $\Lambda^{ab}_{\text{boost}}$ is a 2-form {in local Lorentz space}, in contrast to  the celebrated Iyer-Wald entropy \cite{Wald:1993nt, Iyer:1994ys}, which is the charge of the vector field $\xi_{\text{\tiny{H}}}$ generating a bifurcate Killing horizon. 
\item[III.] $\Lambda^{ab}_{\text{boost}}$ is a part of the non-Abelian local Lorentz symmetry group; therefore, its normalization is rigidly locked. This should be contrasted with the Wald entropy, in which the normalization of $\xi_{\text{\tiny{H}}}$ is not fixed from first principles. 
    \item[IV.] Wald's entropy symmetry generator is $\xi_{\text{\tiny{H}}}$ normalized by the surface gravity $\kappa_{\text{\tiny{H}}}$, i.e. $\frac{2\pi}{\hbar\kappa_{\text{\tiny{H}}}}\xi_{\text{\tiny{H}}}$ \cite{Wald:1993nt, Iyer:1994ys}. Nonetheless, surface gravity is a normalization and observer-dependent quantity. Moreover, $\kappa_{\text{\tiny{H}}}$ depends dynamically on the solution's parameters. This field or parametric dependence ($\delta \kappa_{\text{\tiny{H}}} \neq 0$) can render the associated surface charge non-integrable \cite{Hajian:2015xlp, Hajian:2020dcq}. By contrast, our boost generator \eqref{local-boost-generator} depends only on the invariant binormal, rendering the charge intrinsically integrable.
\end{enumerate}

\vspace*{-4mm}
\section{Derivation of the Second Law}\label{sec:second_law_derivation}
\vspace*{-2mm}

The second law of thermodynamics,  that the entropy of a closed system cannot decrease in time, is the hallmark of any definition of entropy. While our definition of entropy is clearly and explicitly observer-independent, to discuss the second law, one should inevitably choose an observer. The role of the observer in the thermodynamics of a gravitating system is a noted feature, while its many different aspects still deserve to be understood better. Here, we restrict ourselves to the bare minimum. 

As emphasized in the previous section, \eqref{entropy-formula} defines the entropy associated with the codimension-2 region $\Sigma$ as measured by a generic causal observer whose trajectory $\gamma$ lies in the $2d$ plane ${\cal D}$ transverse to $\Sigma$, cf. the geometric setup in section \ref{sec:geometry}, 
\begin{equation}\label{entropy-formula-lambda}
     S[\lambda] = \frac{1}{4 G \hbar} \int_{\Sigma(\lambda)} \d{}^{D-2}x\, \sqrt{q}\, .
\end{equation}
As time evolves, an observer parameterizing their path by $\lambda$ records an entropy variation along the future-oriented causal curve $\gamma$, denoted by $\delta_{\gamma}S[\lambda]$. The second law then requires that:
\begin{equation}\label{statmnt-2ndlaw-chargevar}
\delta_{\gamma}S[\lambda]\geq 0\,.     
\end{equation}

\vspace*{-6mm}
\subsection{Evaluating the charge variation}\label{sec:evaluating_variation}
\vspace*{-2mm}

The extensive literature on the gravitational second law generically works directly with the codimension-2 integrals defining the entropy (like \eqref{charge-in-D-entropy}) when trying to establish \eqref{statmnt-2ndlaw-chargevar}. Here, we take a slightly different route that highlights the role of the causal boundary. We use Stokes' theorem to rewrite the rate of change of the entropy as a codimension-1 integral over the causal hypersurface $\Gamma$:
\begin{equation}\label{2nd-law-chargvar-stokes}
    \begin{split}
   \hspace*{-4.5mm}     \delta_{\gamma}S[\lambda] & = \frac{1}{4G\hbar}\int_{\Sigma(\lambda)} \d{}^{D-2}x\, {\cal L}_{v_\lambda}{\sqrt{q}}  \\
        & =  \frac{1}{4G\hbar}\left[\int_{\Sigma(\lambda)} \d{}^{D-2}x{\cal L}_{v_\lambda}{\sqrt{q}} - \int_{\Sigma(\infty)}\d{}^{D-2}x{\cal L}_{v_\lambda}{\sqrt{q}}\right] \\
        & = - \frac{1}{4G\hbar}\int_{\Gamma} \d{}^{D-2}x\, \d{}\lambda\, \ {\cal L}_{v_\lambda}(  {\cal L}_{v_\lambda}{\sqrt{q}})\, .
    \end{split}
\end{equation}
Notice that $\Gamma$ effectively acts as a causal boundary (similar to constructions in \cite{Adami:2022ktn}). {In the second equality of \eqref{2nd-law-chargvar-stokes}, we assumed that the system eventually settles down to a stationary equilibrium state, $\mathcal{L}_{v_{\lambda}} \sqrt{q}[\lambda\to \infty]=0$ (cf. Fig.~\ref{fig:setup-2}). Similar asymptotic stationarity assumptions are standard in the existing proofs of the second law; see, e.g., \cite{Bousso:2015qqa, Bousso:2015mqa}.}

{To evaluate \eqref{2nd-law-chargvar-stokes} explicitly, we substitute the tangent vector $v^\mu_\lambda$ from \eqref{causal-vector-RB}. The entropy variation is then computed by expanding the nested Lie derivative on the volume form:}
\begin{align}\label{apdxB:chrg-var-CPSF-entropy-timelk-nul}
    \mathcal{L}_{v_{\lambda}}\sqrt{q}=\frac{1}{\sqrt{2}}\mathscr{R} (\mathscr{B}\theta_{l}+\mathscr{B}^{-1}\theta_{n})\sqrt{q}\,,
\end{align}
which yields:
\begin{align}\label{delta-S-R-B}
  \hspace*{-3mm}  \delta_{\gamma}S[\lambda] &= -\frac{1}{8G\hbar}\int_{\Gamma} \d{}^{D-2}x\, \d{}\lambda \sqrt{q}\, \mathscr{R}\,\bigg[  \mathscr{B}^{2}(l^{\mu}\nabla_{\mu} \theta_{l}+\theta_{l}^{2}) \nonumber\\
    & +  \mathscr{B}^{-2}\,(n^{\mu}\nabla_{\mu} \theta_{n} +\theta_{n}^{2}) + \big(l^{\mu}\nabla_{\mu}\theta_{n} +  n^{\mu} \nabla_{\mu} \theta_{l}  + 2\theta_{l}\theta_{n}\big) \nonumber\\ 
    & +\big(\theta_{l}\, n^{\mu}  +  \theta_{n}\, l^{\mu} +  \mathscr{B}^{2}\theta_{l}\,l^{\mu} +  \mathscr{B}^{-2}\theta_{n}\,n^{\mu}\big)\nabla_{\mu}\ln\mathscr{R} \nonumber\\ 
    & +\big(\mathscr{B}^{2}\theta_l l^\mu-\mathscr{B}^{-2} \theta_n n^\mu+ (\theta_l n^\mu-\theta_n l^\mu) \big)\nabla_{\mu}\ln\mathscr{B}\bigg]\,.
\end{align}
{Equation \eqref{delta-S-R-B} is the exact expression for the entropy variation along the causal boundary $\Gamma$. At this stage, the expression contains explicit dependence on the observer's local parametrization $\mathscr{R}$ and boost $\mathscr{B}$, as well as unsimplified null derivative terms. In the following subsection, we will systematically eliminate these redundancies by fixing our coordinate choices and invoking the Raychaudhuri equations.}

\vspace*{-6mm}
\subsection{Gauge fixing and Raychaudhuri equations}\label{sec:gauge_fixing}
\vspace*{-2mm}

Eq.~\eqref{delta-S-R-B} provides the raw variation, which can now be simplified to extract the underlying physics. To this end, we 
\begin{enumerate}[leftmargin=4mm, itemsep=0pt]
\item[A.] invoke geometric identities—specifically the generalizations of the Raychaudhuri equations—and 
\item[B.] appropriately fix the  parameterization and boost functions $\mathscr{B}$ and $\mathscr{R}$.
\end{enumerate}
For the geometric identities, we recall the standard Raychaudhuri equations for null congruences \cite{Gourgoulhon:2005ng}:
\begin{subequations}\label{Raych-Eqs}
\begin{align}
    l^{\mu}\nabla_{\mu} \theta_{l} &= -\frac{1}{D-2} \theta_l^2+ \kappa_l \theta_l - N_l^2- G_{ll}\,,\label{Raych-Eq-l}\\ 
    n^{\mu}\nabla_{\mu} \theta_{n} &= -\frac{1}{D-2} \theta_n^2- \kappa_n \theta_n - N_n^2- G_{nn}\,,\\ 
    n^{\mu}\nabla_{\mu} \theta_{l} &+ l^{\mu}\nabla_{\mu} \theta_{n} = - \frac{2}{D-2}\theta_{n}\theta_{l}+\kappa_{n} \theta_{l}-\kappa_{l}\,\theta_{n} \nonumber \\ 
    &\quad -2N_l\cdot N_n + 2R_{lnln}-2G_{ln}+R\,,
\end{align}
\end{subequations}
where $G_{\mu\nu}=R_{\mu\nu}-\frac{1}{2} R g_{\mu\nu}$ is the $D$-dimensional Einstein tensor, $R_{\mu\nu\alpha\beta}$ is the Riemann curvature tensor, and the relevant projections are defined as:
\begin{equation}
\begin{split}
\hspace*{-4mm}    G_{ll} = & G_{\mu\nu} l^\mu l^\nu\, , \quad G_{nn} = G_{\mu\nu} n^\mu n^\nu\, , \quad G_{ln} = G_{\mu\nu} l^\mu n^\nu\, , \\ 
\hspace*{-2mm}    R_{lnln} &= R_{\mu\nu\alpha\beta}l^\mu n^\nu l^\alpha n^\beta = R_{\mu\nu\rho\sigma} s^\mu v^{\nu} s^{\rho} v^{\sigma} := R_{svsv}\,.
\end{split}
\end{equation}

Next, we specify the functions $\mathscr{B}$ and $\mathscr{R}$. Motivated by the structure of \eqref{delta-S-R-B}, we impose the conditions:
\begin{subequations}\label{observer-choise}
\begin{align}
    v \cdot\nabla\ln\mathscr{B} & \mathrel{\hat{=}} - \frac{1}{\sqrt{2}}\left(\mathscr{B}\kappa_{l}+\mathscr{B}^{-1}\kappa_{n}\right) \, , \label{ansatz-B-R-2}\\
    v\cdot\nabla\ln\mathscr{R} & \mathrel{\hat{=}} - \theta_v  \, .\label{ansatz-B-R}
\end{align}
\end{subequations}
Let us elucidate the physical significance of these choices. Recalling  \eqref{kappa-v}, condition \eqref{ansatz-B-R-2} implies $\kappa_v \mathrel{\hat{=}} 0$, meaning that the observer's velocity vector $v^\mu$ traces out an affinely parameterized geodesic:
\begin{equation}\label{v-geod}
    v\cdot \nabla v \mathrel{\hat{=}} 0\, .
\end{equation}
On the other hand, condition \eqref{ansatz-B-R} implies:
\begin{equation}\label{zeroth-law}
    v\cdot \nabla \ln( \sqrt{q}\, \mathscr{R})  \mathrel{\hat{=}} 0\, ,
\end{equation}
{which states that the product $\sqrt{q}\,\mathscr{R}$ is conserved along the geodesic flow. We note that, as the integrals in \eqref{delta-S-R-B} and \eqref{delta-S-main} show, $\sqrt{q}\,\mathscr{R}$ is the square root of the determinant of the induced metric on $\Gamma$ (recall that $\Gamma$ is the codimension-1 causal surface orthogonal to $s^\mu$, with $s^2=1$, as established in \eqref{vec-s}). Therefore, condition \eqref{zeroth-law} guarantees that the volume form of $\Gamma$ is conserved along the observer's trajectory $\gamma$. Remarkably, these two conditions \eqref{observer-choise} can be compactly packaged into a single geometric equation for the observer's velocity:}
\begin{equation}\label{2nd-law-observer}
   \inbox{ v_{\lambda} \cdot \nabla v_\lambda = -\theta_{v_{\lambda}} v_{\lambda}\, .}
\end{equation}
Here $\theta_{v_{\lambda}} := q_{\mu\nu} \nabla^{\mu}v_{\lambda}^{\nu} = \mathscr{R}\, \theta_{v}$.

With these gauge choices implemented, substituting \eqref{observer-choise} and the Raychaudhuri equations back into the variation \eqref{delta-S-R-B} triggers massive algebraic cancellations. The resulting compact expression for the entropy variation becomes:
\begin{equation}\label{delta-S-main-0}
  \begin{aligned}
    \delta_{\gamma}S[\lambda] 
    = \frac{1}{4G\hbar}\int_{\Gamma} \mathrm{d}^{D-2}x\ \mathrm{d}\lambda\ \sqrt{q}\ \mathscr{R} 
   \bigg[ \frac{\theta_{v}^{2}}{D-2} + N_v^2+ R_{vv} - R_{svsv} \bigg]\, .\nonumber 
  \end{aligned}
\end{equation}
Here $R_{vv} := R_{\mu\nu} v^\mu v^\nu$. This expression simplifies even further, as the contribution of the tidal term $R_{svsv}$ vanishes. To see this, consider the geodesic deviation equation for two nearby causal geodesics (such as neighboring points along our observer path $\gamma$) separated by the spacelike unit normal vector $s^\mu$:
\begin{equation}\label{geo-devi}
    s_\mu \frac{D^2 s^\mu}{\d{}\lambda^2}= R_{\mu\nu\rho\sigma} s^\mu v^{\nu} v^{\rho} s^{\sigma}=- R_{svsv}\,.
\end{equation}  
Integrating this expression by parts along the geodesic $\gamma$ yields:
\begin{equation}\label{int-Rsvsv}
\begin{split}
 \int_\gamma \d{}\lambda\, R_{svsv} = -\int_\gamma \d{}\lambda\, \left(s_\mu \frac{D^2 s^\mu}{\d{}\lambda^2}\right)  = \int_\gamma \d{}\lambda \left(\frac{\d{}s^\mu}{\d{}\lambda}\right)^2\,.
\end{split}
\end{equation}
Eq.~\eqref{cov-s-decompo} together with $\kappa_v \mathrel{\hat{=}} 0$ imply that:
\begin{equation}
     \frac{d s^\mu}{\d{}\lambda} = v\cdot \nabla s^{\mu} \mathrel{\hat{=}} 0\, .
\end{equation}
{Combining this result with \eqref{int-Rsvsv} yields $\int_\gamma \d{}\lambda\, R_{svsv} = 0$. Consequently, we arrive at our final master formula for the entropy variation:}
\begin{equation}\label{delta-S-main}
  \begin{aligned}
   \inbox{\hspace{-.4 cm} \delta_{\gamma}S[\lambda] 
    = \frac{1}{4G\hbar}\int_{\Gamma} \mathrm{d}^{D-2}x\mathrm{d}\lambda \sqrt{q}\mathscr{R} \bigg[ \frac{\theta_{v}^{2}}{D-2} + N_v^2+ R_{vv} \bigg]\hspace{-.4 cm}}
  \end{aligned}
\end{equation}

\vspace*{-6mm}
\subsection{Energy conditions and the positivity of entropy change}\label{sec:energy_conditions}
\vspace*{-2mm}

Equation \eqref{delta-S-main} represents our main result. To establish the second law $\delta_\gamma S[\lambda] \ge 0$, we examine the sign of the individual terms in the integrand. The expansion-squared term is manifestly positive-definite. The shear term $N_v^2$ is also positive-definite because, by definition, both indices of $N^v_{\mu\nu}$ lie strictly within the spacelike tangent space of $\Sigma$. Consequently, the only term requiring detailed analysis is $R_{vv}$.  

To determine the sign of the $R_{vv}$ term, we invoke the gravitational field equations. Expressing the Ricci tensor in terms of the energy-momentum tensor via the Einstein field equations, $R_{\mu\nu} - \frac{1}{2}R g_{\mu\nu} = 8\pi G T_{\mu\nu}$, we obtain:
\begin{equation}\label{Rvv-SEC}
    R_{vv} = R_{\mu\nu}v^\mu v^\nu = 8\pi G \left(T_{\mu\nu}-\frac{T}{D-2} g_{\mu\nu}\right) v^\mu v^\nu\,.
\end{equation} 
The requirement that $R_{vv} \geq 0$ for any future-oriented timelike vector $v^\mu$ entails the strong energy condition (SEC) \cite{Ford:1994bj, Curiel:2014zba, Wald:1984rg, Kontou:2020bta}. In fact, for the second law, we have the integrated form of the SEC along the causal history of the bounding surface. If the matter content satisfies an integrated SEC along the segments of the causal curve $\gamma$, then each term in  \eqref{delta-S-main} is non-negative, establishing the classical gravitational second law. $\Box$

We close this part with the comment that while the second law is valid for all observers following geodesics in parameterization adapted in \eqref{2nd-law-observer}, the exact expression of $\delta_\gamma S$ depends on the gauge-fixings and one may have other positive-definite contributions in the right-hand side of \eqref{delta-S-main}. A slightly weaker gauge fixing may be found in \cite{To-appear-2ndlaw-observer}, where one requires just the hypersurface orthogonality of the causal vector $v$ (and not that of null vector fields $l,n$).

\vspace*{-6mm}
\subsection{Null limit}\label{sec:null-limit}
\vspace*{-2mm}
So far the analysis was for a generic causal, timelike or null, observer. The specific case of null observers is more closely related to the cases discussed and analyzed in the literature. So, to more directly compare with the existing literature, it is instructive to specialize our analyses for the null case. There are two ways of doing so: repeating the whole analyses for the $v^2=0$ case, or taking the null limit. The latter may be obtained via an appropriate double scaling limit, $\mathscr{B}\to\infty$ and $\mathscr{R}\to 0$. One may check that these two, as expected by physical continuity, yield the same final result. Here we show the latter. Starting from \eqref{delta-S-R-B} suggests to define the double scaling limit as $\mathscr{B}\to\infty$ and $\mathscr{R}\to 0$, while keeping $\tilde{\mathscr{R}}:=\mathscr{B}^2 \mathscr{R}$ fixed. Applying this limit to \eqref{delta-S-R-B}, we obtain
\begin{equation}\label{delta-S-R2B}
\begin{split}
  \hspace*{-3mm}  \delta_{\gamma}S[\lambda] = -\frac{1}{8G\hbar}\int_{\Gamma} & \d{}^{D-2}x\, \d{}\lambda \sqrt{q}\, \tilde{\mathscr{R}}\,\bigg[  (l^{\mu}\nabla_{\mu} \theta_{l}+\theta_{l}^{2}) \\
    & +
    \theta_{l} 
    \big(l \cdot \nabla \ln{\mathscr{R}} + l \cdot \nabla \ln\mathscr{B}\big)\bigg]\,.    
\end{split}
\end{equation}
Next, we note that in the double scale limit, we are still free to choose derivatives of $\ln\mathscr{R}, \ln\mathscr{B}$. We do so, by applying the limit to \eqref{v-geod} and \eqref{zeroth-law}, which recalling \eqref{CD-l-inde}, reduce to, 
\begin{equation}\label{null-observer}
   l\cdot \nabla \ln\mathscr{B} = -\kappa_l\, , \qquad l\cdot \nabla\ln(\sqrt{q} \mathscr{R})=0\,.
\end{equation}
That is, $l_\mu$ is along a non-affinely parametrized null geodesic. Substituting the null Raychaudhuri equation \eqref{Raych-Eq-l} into \eqref{delta-S-R2B}, we find that the variation simplifies to:
\begin{equation}
\delta_{\gamma}S[\lambda]  
     = \frac{1}{8G\hbar}\int_{\Gamma} \d{}^{D-2}x \d{}\lambda \sqrt{q}\tilde{\mathscr{R}}\bigg( \frac{\theta_l^2}{D-2} + N_l^2+ R_{ll} \bigg).  
\end{equation}
{In the null limit $\mathscr{B}\ l^\mu$ is an affinely parametrized null geodesic}, the Ricci term $R_{ll}$ is  non-negative as implied by NEC ($T_{\mu\nu}l^\mu l^\nu \geq 0$) and the Einstein field equations. This guarantees that $\delta_{\gamma}S \geq 0$, which is consistent with the standard results established for null horizons \cite{Wall:2009wm}.

\vspace*{-4mm}
\section{Discussion and Outlook}\label{sec:discussion}

In this work, we revisited the gravitational second law from a generalized perspective. By defining the gravitational entropy of a spatial region enclosed by a codimension-2 surface $\Sigma$ as the surface charge associated with local boosts, we completely disentangled the definition of entropy from specific spacetime structures like event horizons or light sheets. Using the CPSF, we ensured that this charge is integrable and mathematically robust. We then evaluated its variation along a generic causal curve $\gamma$, maintaining full general covariance throughout the procedure, and showed that the second law holds for a generic causal free-fall observer whose trajectory is parametrized according to the ``expansion-adjusted'' prescription \eqref{2nd-law-observer}, provided that the SEC is satisfied in an integrated sense along the observer's path.

\textbf{Observer dependence of thermodynamic description of gravitating systems.}  
While our analyses here are classical, thermodynamics description of gravitating systems is better stated and understood in the semiclassical regime. In this regard, Unruh famously showed that the extension of Einstein's equivalence principle to the semiclassical regime implies that an accelerated observer finds herself in a thermal bath while an inertial observer provides a non-thermal description \cite{Unruh:1976db, Grumiller:2022qhx}. Similarly, a free-fall observer near the horizon of a black hole \textit{locally} provides a non-thermal description, while a non-free-fall (fiducial) observer gives the standard Hawking's thermal description \cite{Susskind:2005, Grumiller:2022qhx}. On the other hand, Jacobson's view of ``gravity = spacetime thermodynamics'' \cite{Jacobson:1995ab}, reinforced through surface charge analyses and CPSF, and made explicit in the current work as well as in our recent papers \cite{Shajiee:2026coz, To-appear-2ndlaw-observer}, suggests that there exist well-defined thermodynamical notions in generic gravitating systems. So, there seem to be two different thermal features, one of which is available to accelerating, non-free-fall observers and the other to free-fall observers. 
Investigating and understanding the relation between the two would shed light not only on the gravity = spacetime thermodynamics proposal, but on the black hole information problem.

\textbf{Dynamical Entropy.}  
In this paper, we derived gravitational entropy as a particular local Lorentz (boost) charge, which coincides with the area of a codimension-2 cross section for Einstein gravity theory. Based on spacetime diffeomorphisms, Hollands, Wald, and Zhang (HWZ) \cite{Hollands:2024vbe} proposed a different expression for gravitational entropy in systems close to stationarity and involving nearly stationary black holes, in the perturbative dynamical regime (see also \cite{Visser:2024pwz, Rignon-Bret:2023fjq}). In our recent work, we extended this dynamical entropy to a Noether charge associated with generic null surfaces in arbitrary dynamical backgrounds \cite{Shajiee:2026coz}. It would be interesting to investigate whether this dynamical entropy can likewise be derived within the first-order formulation as a charge associated with the internal local Lorentz symmetry. Furthermore, understanding the observer dependence of dynamical entropy would provide another interesting direction for future research.

\textbf{Energy Conditions and the Null Limit.} 
We found that the classical gravitational second law ($\delta_{\gamma}S \geq 0$) is satisfied locally provided the matter content obeys the integrated SEC over any segment of the observer's curve. Geometrically, the SEC implies that a congruence of causal geodesics is convergent in a sufficiently small neighborhood \cite{Curiel:2014zba}, a feature famously utilized in Hawking's singularity theorem \cite{Hawking:1966vg}. Our entropy is defined as an integral over the surfaces pierced by these geodesics. Therefore, it is natural that the SEC dictates its monotonic behavior.

However, the SEC is known to be violated in physically relevant scenarios, most notably during cosmic inflation or in our current accelerated expanding universe driven by dark energy \cite{Mukhanov:2005sc}. In such de Sitter-like phases, the universe is bounded by a cosmological horizon to which one associates the Gibbons-Hawking entropy \cite{Gibbons:1977mu}. Our current analysis focused strictly on deriving the local version of the second law. It is highly plausible that including global structures (like cosmological horizons) and their associated entropy contributions could relax the SEC requirement down to the NEC, which is standard in many generalized second law proofs \cite{Wall:2009wm, Bousso:2015eda}. 

\textbf{Thermodynamics of Spacetime.} 
{Our approach introduces an integral over a causal codimension-1 hypersurface $\Gamma$, which acts as the causal boundary of the observer's history. Viewed through this lens, our main result \eqref{delta-S-main} strongly resembles an expression of the Clausius law (energy conservation) $T\delta S = \delta E + P\delta V$. In this thermodynamic interpretation, the left-hand side is the $T\delta S$ term, $N_v^2$ represents the gravitational kinetic energy (or dissipation) and $R_{vv}$ is the gravitational work done by the matter ($P_{\text{eff}}\delta V$).}

This suggests a deep connection to Jacobson's derivation of the Einstein field equations from the local thermodynamics of null surfaces \cite{Jacobson:1995ab} (see also the ``null surface thermodynamics'' proposal in \cite{Adami:2021kvx}), and opens a pathway to extend Jacobson's ideas to generic causal surfaces. Because the CPSF evaluates charges purely on-shell, our entropy formula inherently satisfies local thermodynamics as a consequence of the equations of motion. Reversing this logic, one could postulate  \eqref{delta-S-main} as a fundamental thermodynamic axiom for \textit{timelike} surfaces, from which the gravitational field equations are expected to emerge. 

\textbf{Beyond Einstein Gravity.} 
Finally, while this paper focused on the Einstein-Hilbert action, our foundational definition of entropy as the CPSF charge of local boosts applies universally to any diffeomorphism-invariant higher-curvature theory (e.g., Lovelock or $f(R)$ gravity). The integration strategies presented here will carry over, provided one utilizes the modified Raychaudhuri equations \eqref{Raych-Eqs} and identifies the appropriate generalized energy conditions necessary to enforce the positivity of the entropy variation \cite{progress-2ndlaw}.

\vskip 2mm
\textbf{Acknowledgment.} 
We thank Hamed Adami for valuable discussions and collaboration at the early stages of this work. We also thank Kamal Hajian and Delong Kong for their helpful comments on the draft. M.M.Sh.J. acknowledges the hospitality of the Beijing Institute of Mathematical Sciences and Applications (BIMSA). V.R.S. is supported by the Iran National Science Foundation (INSF) under project No. 4031681. M.M.Sh.J. and V.R.Sh. acknowledge the support of INSF research chair No. 4045163.

\appendix

\vspace*{-7mm}
\section{A quick overview on the second law literature}\label{apdx:overview}
\vskip -3mm

The literature on the second law of thermodynamics in gravitational systems is very vast. In the introduction of the main text, we only reviewed the part that was immediately relevant to our derivations and main results. Here, we provide a slightly more complete overview (again, not very detailed or comprehensive) of the literature. We focus on the main conceptual or technical advances on the topic. 

Historically, the connection between gravitational systems and thermodynamics started in the early 1970s and in the context of black holes. The hallmarks of the early developments are (1) Bekenstein's entropy and entropy bound \cite{Bekenstein:1972tm, Bekenstein:1973ur}; (2) the Bardeen-Carter-Hawking paper \cite{Bardeen:1973gs}; (3) Hawking's area theorem \cite{Hawking:1971vc, Hawking:1974rv}. This led to the area law for black hole entropy and the notion of ``generalized second law'' stating that the total entropy of a system consisting of black holes plus other matter does not decrease in time. Viewed in light of (Einstein's) equivalence principle, this meant that thermodynamic behavior and features should not remain limited to black holes, and even kinematic effects, like dealing with accelerated observers, should give rise to thermodynamic features, e.g. the Unruh effect \cite{Unruh:1976db}. 

The next wave of conceptual advancements on the topic came in the 1990s and from the idea of connecting the universality of gravity (manifested through Einstein's GR) and the universality of thermodynamics. This had two important implications/outcomes: (1) Thermodynamical features should not be limited to Einstein's GR and should be generic features of any generally covariant theory. The first concrete steps towards this were taken by Wald \cite{Wald:1993nt, Iyer:1994ys}; see also \cite{Jacobson:1993xs, Jacobson:1993vj, Hayward:1993wb}. (2) As Jacobson famously showed, one can derive Einstein's equations from the (local) first law of thermodynamics on a null surface \cite{Jacobson:1995ab}. 

Any definition of entropy must satisfy some version of the second law. The first work in this direction goes back to Hawking's area theorem \cite{Hawking:1971vc, Hawking:1971tu}, where the second law was derived as a consequence of the weak energy condition for the matter. The area theorem applies to event horizons, which are only idealized theoretical notions, not detected in nature (especially once the quantum effects and Hawking radiation are also taken into account). Hayward \cite{Hayward:1993wb} extended the derivation of the laws of thermodynamics to apparent and trapping surfaces (horizons). He argued that ``the area of a future trapping horizon is increasing, constant or decreasing if the horizon is spatial, null or Lorentzian, respectively.'' Hayward's work is based on the future trapping horizon (FTH), in which each corner of it is a marginally trapped codimension-2 hypersurface with respect to the expansion of the generating vector field. FTH may be perceived as a generalization of the dynamical horizon in which the expansion of one of the null normal vectors vanishes, and the expansion of the other null normal is negative everywhere; see \cite{Bousso:2015mqa, Bousso:2015qqa} for a more recent account. Ashtekar-Krishnan verified the first and second laws for such horizons in full, non-linear general relativity~\cite{Ashtekar:2002ag, Ashtekar:2003hk}.

The notion and idea of entropy bounds are closely related to the second law, as was initially put forward by Bekenstein \cite{Bekenstein:1972tm, Bekenstein:1973ur}. Bousso promoted the Bekenstein bound to the covariant entropy bound \cite{Bousso:1999xy}, where it was conjectured that the entropy of a hypersurface generated by surface-orthogonal null geodesics with non-positive expansion bounded by a two-dimensional surface does not exceed $A/4$, where $A$ is the area of the two-dimensional surface; see \cite{Flanagan:1999jp, Bousso:2014sda} for proofs. He also argued that the conjecture is to be valid in all spacetimes admitted by Einstein's equation. He then introduced the holographic screens using these codimension-2 spacelike hypersurfaces \cite{Bousso:1999cb}. Later, Bousso and Engelhardt showed that the area of the future holographic screen increases monotonically along its foliation by marginally trapped surfaces---a ``new area law'' that can be thermodynamically interpreted as a second law by the Bousso bound~\cite{Bousso:2015mqa, Bousso:2015qqa}.

Jacobson-Kang-Myers~\cite{Jacobson:1994qe, Jacobson:1995uq} considered higher curvature gravity and proved the second law by defining the change in entropy along the null congruence generating the event horizon under any dynamical evolution as a codimension-2 integral of expansion over a compact spacelike cross section of the horizon. They used the cosmic censorship and the null energy condition on the matter to conclude that the entropy can never decrease. In addition, in~\cite{Jacobson:1993vj}, they noted ambiguities in the definition of entropy as a Noether charge for compact nonstationary horizons, and asserted that the preferred definition is the one that obeys the second law.

Aron Wall has made a notable contribution to the verification of the second law of black hole thermodynamics in a generally covariant theory of gravity, e.g. in \cite{Wall:2015raa} (see also \cite{Rignon-Bret:2023fjq}). He showed that ``a second law indeed exists in all such theories of higher curvature gravity. This holds true provided one only considers linearized perturbations of the gravitational fields which may be sourced by a first-order perturbation to the energy-momentum tensor, evaluated on a stationary black hole background'' (or, more precisely, a bifurcate Killing horizon). Crucially, Wall's work depends on the first law to derive the second law. Wall focuses on the cases with matter and does not discuss the pure gravity case in detail.

Energy conditions are crucial ingredients going into discussions of the (generalized) second law. Energy conditions, which are typically related to positivity of certain combinations of the energy-momentum tensor, impose restrictions on the evolution of spacetime via Einstein's equations \cite{Ford:1994bj, Curiel:2014zba}. It is known that dominant or strong energy conditions may be violated in classical field theories, e.g., those used in inflationary models, while they typically respect weak and null energy conditions \cite{Kontou:2020bta}. Moreover, non-minimally coupled classical fields or higher derivative gravity theories may also exhibit violation of energy conditions \cite{Flanagan:1996gw}. However, quantum mechanically, field theories respect weaker energy conditions, and averaged or quantum energy conditions are enough to ensure physical consistency or desired features of the theory or the spacetime, e.g., \cite{Ford:1978qya, Wald:1991xn, Graham:2007va, Kontou:2015yha, Fewster:2018pey, Kontou:2020bta, Kontou:2023ntd}. The null energy condition (NEC) or its averaged version is the weakest energy condition and is usually viewed as the requirement for causality and the quantum stability of quantum field theories \cite{Faulkner:2016mzt, Hartman:2016lgu}. (Averaged) NEC is argued to be a necessary condition for having the second law for event Killing horizons \cite{Wall:2009wi, Wall:2010jtc, Wall:2011hj}. 

We close the overview with two comments: (1) Ref.~\cite{Neiman:2013taa} also discusses a boost/corner viewpoint in GR and Lovelock theories. As a result of gluing across a codimension-2 corner, the gravitational action acquires a non-additive imaginary part controlled by the (Wald) entropy of the surface, which admits a natural entanglement-entropy interpretation, because the corner angle is a Lorentzian boost. Whereas here we formulate gravitational entropy for generic codimension-2 surfaces as the surface charge associated with the local boosts, placing a similar boost-centered structure into a charge/flux framework that we use to establish a classical second-law statement. (2) There are numerous other valuable works on the second law and related areas; for some recent work, see~\cite{ Hollands:2022fkn, Bhattacharyya:2021jhr, Wang:2021zyt, Biswas:2022grc, Lin:2022ndf, Dhivakar:2023mai, Davies:2023qaa, Wall:2024lbd, Hollands:2024vbe}.


\bibliographystyle{fullsort.bst}
\bibliography{reference}
\end{document}